\documentclass{article}
\usepackage{color}
 \usepackage{amsbsy}
 \usepackage{amsfonts}
\usepackage{amssymb}
\usepackage{amsmath}
\usepackage{graphics}
\usepackage{graphicx}

\begin{document}

\title{In-medium  covariant propagator of baryons under a strong magnetic field: effect of the intrinsic magnetic moments}
\author{R. M. Aguirre and A. L. De Paoli.\\
\it{Departamento de Fisica, Facultad de Ciencias Exactas}, \\\it{Universidad Nacional de La Plata,} \\
\it{and IFLP, UNLP, CONICET. Argentina.}}

\date{}

\maketitle

\begin{abstract}
We obtain the covariant propagator at finite temperature for
interacting baryons immersed in a strong magnetic field. The
effect of the intrinsic magnetic moments on the Green function are
fully taken into account. We make an expansion in terms of
eigenfunctions of a Dirac field, which leads us to a compact form
of its propagator. We present some simple applications of these
propagators, where the statistical averages of nuclear currents
and energy density are evaluated.


\end{abstract}

\newpage

\section{Introduction}

 The dynamics of matter subject to strong magnetic
fields has been widely studied in the past \cite{LAI}, and it has
received renewed interest due to the analysis of different
experimental situations. For instance, some investigations of the
last decade\cite{KHARZEEV,KHARZEEV2,MO,SKOKOV} have pointed out that matter
created in heavy ion collisions could be subject to very intense
magnetic fields. As a consequence the particle production  can
exhibit a distinguishable anisotropy. A preferential emission of
charged particles along the direction of the magnetic field is
predicted in~\cite{KHARZEEV,KHARZEEV2} for noncentral heavy ion collisions,
due to magnetic intensities  $e\, B \sim 10^2$ MeV$^2$. Improved
calculations taking care of the mass distribution of the colliding
ions \cite{MO}, does not modify essentially the magnitude of the
produced fields.  Furthermore, the numerical simulations performed
by \cite{SKOKOV} predict larger values $e\,
B \sim m_\pi^2 \sim 2 \times 10^4$ MeV$^2$.\\
In a very different scenario, the presence of strong magnetic
fields is the key issue that distinguishes a kind of astronomical
compact objects. The analysis of the observational data in the
range from soft X to soft gamma radiation, has showed the features
of a class of neutron stars named Soft Gamma Repeaters and
Anomalous X Ray Pulsars. These isolated stars are characterized by
a sustained X-ray luminosity with energy in the soft (0.5-10 keV)
or hard (50-200 keV) spectrum. They can show a time variability,
with pulsations at relatively long spin periods. In particular,
the Soft Gamma Repeaters exhibit a bursting activity which
includes giant flares as a rare manifestation. Both cases can be
described within the magnetar model
\cite{DUNCAN,THOMPSON,THOMPSON2}, where the X-ray emission as well
as bursting are attributed to the dissipation and decay of very
strong magnetic fields. Their intensity has been estimated around
$10^{15}$ G at the star surface, and could reach much higher
values in the dense interior of the star. The availability of an
increasing amount of precision data opens the question on how well
the current theoretical description of nuclear matter can fit this
empirical evidence.

The properties of the dense hadronic medium have been properly
described within a covariant model of the hadronic interaction
known as Quantum Hadro-Dynamics (QHD) \cite{SW}. It has been used
to study the structure of neutron stars and particularly to
analyze hadronic matter in the presence of an external magnetic
field \cite{CHAKRABARTY,
BRODERICK,MALLICK,DONG,DONG2,RHABI,REZAEI}. The versatility of
this formulation allows the inclusion of the intrinsic magnetic
moments in a covariant way. Due to the strength of the
baryon-meson couplings, the mean field approximation (MFA) is
usually employed. Within this approach the meson fields are
replaced by their expectation values and assimilated to a
quasi-particle picture of the baryons. Finally the meson mean
values are obtained by solving the classical meson equations
taking as sources the baryonic currents. This scheme is
conceptually clear and easy to implement, however it is not
evident at all how to include further
corrections if they were needed.\\
In recent years several publications have stressed the
role of the intrinsic magnetic moments on the statistical properties
of hadronic systems such as the matter susceptibility and
magnetization \cite{DONG,DONG2, RHABI}, the rise in the population of
hyperons in stellar matter \cite{DONG,DONG2}, and the saturation
properties of nuclear matter \cite{REZAEI}. The variation of the
magnetic moments of hadrons within the nuclear environment has been
pointed out in recent investigations \cite{RYU}. \\

The purpose of this work is twofold. In first place we construct the
covariant propagator of fermions, both neutral and charged, in the
presence of an external magnetic field. We give a full treatment
including their intrinsic magnetic moments. \\
Expressions for the covariant propagator of a charged particle
subject to an external magnetic field have been presented long
time ago \cite{SCHWINGER,RITUS}, and this is a subject of
continuous development \cite{KUZNETSOV,MIRAN}. However, the
effects of the magnetic moment have been neglected assuming its
smallness.
 Exceptionally, in Ref. \cite{CHEOUN} the proton
propagator in-vacuum has been presented.\\
For magnetic intensities greater than $5\times 10^{17}$ G, the
influence of the magnetic moments must be taken into account in
the determination of the stable configuration of matter
\cite{A&B}, and the evaluation of thermodynamical
properties \cite{DONG,DONG2,RHABI,REZAEI}. \\
On the other hand, we give here an extension of the
Dirac field propagator appropriate to include density
and temperature effects in the study of hadronic
systems subject to very strong magnetic fields. It is shown that the
mean values of the particle densities and currents,
agree with the results obtained for nuclear matter
within the QHD model in the MFA.
\\
We present a detailed derivation, using a clearly stated notation.
Our results open the possibility of using the diagrammatic
techniques of the field theory to study quantum
corrections and statistical averages of physical processes
developing under strong magnetism, including the effects of the
magnetic moments of baryons. Thus we propose a complementary tool
to extend the analysis of related investigations \cite{AYALA,DESA,CHEOUN}.\\

The organization of this work is as follows. In the next section
we summarize the classical solutions for a Dirac
field in the presence of an external magnetic field, considering
the intrinsic magnetic moments. For this purpose we follow the
general guidelines of \cite{BRODERICK}. This complete set of
solutions is used to make an expansion of the quantum fields,
including the appropriate measure of integration in the phase
space. Following the standard prescriptions we evaluate the
in-medium nucleon propagator in sect. \ref{GreenP&N}. These results
are interpreted within the context of the QHD model, and we
evaluate nucleon densities and energy densities in sect. \ref{QHD}.
Finally, in the last section we present a summary of our results.

\section{Dirac solutions for nucleons with magnetic moment}

The Lagrangian density for Dirac particles of mass $m_b$ , with
anomalous magnetic moments $\kappa_b$, interacting through scalar
$\sigma$ and vector $\omega$ mesons, and under the influence of an
external electromagnetic field $A$, is given by (~$\hbar$~=~1~,
$c~=~1$~) \cite{ITZYKSON}
\begin{eqnarray}
\mathcal{L}&=&\sum_b \,\bar{\Psi}^{(b)}\left[\gamma_\mu(i\,\partial^\mu-q_b\,A^\mu- g_\omega\,\omega^\mu) +g_\sigma\,\sigma-m_b-\frac{\kappa_b}{2}\,\sigma^{\mu\nu}\,\mathcal{F}_{\mu\nu}\right]\Psi^{(b)}\nonumber\\
&-&\frac{1}{4}\,\mathcal{F}_{\mu\nu}\,\mathcal{F}^{\mu\nu}
+\frac{1}{2}\,(\partial_\mu\sigma\,\partial^\mu\sigma-m_\sigma^2\,\sigma^2)
-\frac{1}{4}\,\Omega_{\mu\nu}\,\Omega^{\mu\nu}+
\frac{1}{2}\,m_\omega^2\,\omega_\mu\,\omega^\mu\label{Lagran}
\end{eqnarray}
where
$\mathcal{F}^{\mu\nu}=\partial^\mu\,A^\nu-\partial^\nu\,A^\mu$ and
$\Omega^{\mu\nu}=\partial^\mu\,\omega^\nu-\partial^\nu\,\omega^\mu$
are the electromagnetic and vector meson field strength tensors,
$q_b$ denotes the electric charge, and $g_{\sigma,\omega}$ the
strong coupling constants, and $\sigma^{\mu\nu}=i/2\,[\gamma^\mu,
\gamma^\nu]$. We study the case of a constant external magnetic
field $B$ applied along the z axis.  In order to fix ideas and
facilitate the comparison with previous results \cite{BRODERICK},
we choose the gauge $A^\mu=(0,0,B x,0)$. To simplify the
discussion, we consider first baryons interacting only with $B$,
and mesons will be included later. \\It must be mentioned that the
remaining content of this section, has been studied long time ago,
see for instance \cite{BRODERICK}. But we present here a summary
in
order to state clearly the notation used.\\
In this approach the classical eigenstates
$\psi^{(b)}=\phi^{(b)}\,e^{-i E^{(b)}\,t}$ of the
Dirac equation satisfy
\begin{equation}
[\vec{\alpha}.\vec{\pi}+\gamma^0\,m_b-i\gamma^0\,\gamma^1\gamma^2\,\kappa_b\,B]\phi^{(b)}=E^{(b)}\,\phi^{(b)}
\label{eigenD} \end{equation} with
$\vec{\alpha}=\gamma^0\,\vec{\gamma}$ and
$\vec{\pi}=-i\vec{\nabla}-q_b\,\vec{A}$. The label $s$ indicates
the alignment of the magnetic moment with the external field. It
must be borne in mind that when considering nuclear particles, we
can write
 $\kappa_b=\chi_b \mu_N$, with the anomalous moments $\chi_p=2.79$ for protons,
$\chi_n=-1.91$ for neutrons, and $\mu_N$ the nuclear magneton.

\subsection{Charged states}\label{P}

The particle solutions for energies $E_{n s}$  are given by
\\$\phi_{n s p_y p_z}^{(+)(p)}(\xi,y,z)= e^{i(p_y y+p_z
z)}\,e^{-\xi^2/2}\,u_{n s p_z}(\xi)$ with
\begin{eqnarray}
u_{n s p_z}(\xi)=
N_{n s} \left(
\begin{array}{c}
H_n(\xi)\\ \\\frac{2\,n\,s\,p_z\,\sqrt{q B}\,i}{(\Delta_n+s\,m_p)\,(E_{n s}+s\,\Delta_n-\kappa_p B)}\,H_{n-1}(\xi)\\ \\
\frac{p_z}{E_{n s}+s\,\Delta_n-\kappa_p B}\, H_n(\xi)\\
\\-\frac{2\,n\,s\,\sqrt{q B}\,i}
{\Delta_n+s\,m_p}\,H_{n-1}(\xi)\\
\end{array}
\right) \label{Spinorp}
\end{eqnarray}
and,
\begin{eqnarray} \xi&=&(-p_y + q B x)/\sqrt{q B} \\ \nonumber \\
\Delta_n&=&\sqrt{m_p^2+2 n q B}\\ \nonumber \\ \nonumber \\E_{n s}&=&\sqrt{p_z^2+(\Delta_n-s\,\kappa_p B)^2}\\ \nonumber \\
N_{n s}^2&=&\frac{\sqrt{q B}}{4\,\sqrt{\pi}(2 \pi)^2\,2^n\,n!}
\frac{(\Delta_n+s\,m_p)\, (E_{n s}+s\,\Delta_n-\kappa_p
B)}{m_p\,(\Delta_n-s\,\kappa_p B)}
\end{eqnarray}
$H_n$ stands for the Hermite polynomials, and $n\geq1$. \\ In the
case $n=0$ the physical eigenstate corresponds to $\phi_{0 p_y
p_z}^{(+)(p)}(\xi,y,z)= e^{i(p_y y+p_z z)}\,e^{-\xi^2/2}\,u_{0
p_z}$, with
\begin{eqnarray}
u_{0 p_z}=
N_0 \left(
\begin{array}{c}
1\\ \\0\\ \\ \frac{p_z}{E_0+m_p-\kappa_p B}\\ \\0\\
\end{array}
\right) \label{Spinorp0}\end{eqnarray} and
\begin{eqnarray}
E_0&=&\sqrt{p_z^2+(m_p-\kappa_p B)^2}\\ \nonumber \\
N_0^2&=&\frac{\sqrt{q B}}{2\,\sqrt{\pi}(2 \pi)^2}
\frac{E_0+m_p-\kappa_p B}{m_p-\kappa_p B}
\end{eqnarray}
Another solution exists for $n=0$ and $s=-1$, with eigenvalue
$\sqrt{p_z^2+(m_p+\kappa_p B)^2}$, but  it
is asymptotically divergent.\\
The antiparticle states $\phi_{n s}^{(-) (p)}$ correspond to the
negative eigenvalues $-E_{n s}$ and have the eigenfunctions
$\phi_{n s p_y p_z}^{(-) (p)}(\xi,y,z)= e^{-i(p_y y+p_z
z)}\,e^{-\eta^2/2}\,v_{n s p_z}(\eta)$ with
\begin{eqnarray}
v_{n s p_z}(\eta)=N_{n s}
\left(
\begin{array}{c}
\frac{p_z}{E_{n s}+s\,\Delta_n-\kappa_p B}\,H_n(\eta)\\
\\ \frac{2\,n\,s\,\sqrt{q B}\,i}
{\Delta_n+s\,m_p}\,H_{n-1}(\eta)\\ \\
H_n(\eta)\\ \\\frac{-2\,n\,s\,p_z\,\sqrt{q B}\,i}{(\Delta_n+s\,m_p)\,(E_{n s} +s\,\Delta_n-\kappa_p B)}\,H_{n-1}(\eta)\\
\end{array}
\right)
\end{eqnarray}
where $\eta=(p_y + q B x)/\sqrt{q B}$ and  $n\geq1$. While for $n=0$ the antiparticle state
$\phi_0^{(-) (p)}$ has negative energy $-E_0$ and its wave
function reads $\phi_{0 p_y p_z}^{(-) (p)}(\eta,y,z)= e^{-i(p_y
y+p_z z)}\,e^{-\eta^2/2}\,v_{0 p_z}$ with
\begin{eqnarray}
v_{0 p_z}=N_0 \left(
\begin{array}{c}
\frac{p_z}{E_0+m_p-\kappa_p B}\\ \\0\\ \\
1\\ \\0\\
\end{array}
\right)
\end{eqnarray}
The eigenstates are normalized according to \cite{BJORKEN}
\begin{eqnarray}
<\bar{\phi}_{n s p'_y p'_z}^{(\pm) (p)}|\phi_{n s p_y p_z}^{(\pm)
(p)}>&=&\pm \delta(p'_y-p_y)\, \delta(p'_z-p_z)
\end{eqnarray}
and therefore satisfy the covariant orthogonal conditions
\begin{eqnarray}
<\phi_{n' s' p'_y p'_z}^{(\pm)(p) \dag}|\phi_{n s p_y p_z}^{(\pm)
(p)}>&=& \frac{E_{n
s}\,\Delta_n}{m_p\,(\Delta_n-s\,\kappa_p\,B)}\,
\delta_{n' n}\,\delta_{s' s}\,\delta(p'_y-p_y)\,\delta(p'_z-p_z)\nonumber \\
<\phi_{n' s' p'_y p'_z}^{(+)(p)\dag }|\phi_{n s p_y p_z}^{(-)
(p)}>&=& <\phi_{n' s' p'_y p'_z}^{(-) (p) \dag}|\phi_{n s p_y
p_z}^{(+)(p)}>=0 \label{CPROTON}\end{eqnarray} These conditions
also include the case $n=0, s=1$, if $\Delta_0=m_p$ is assumed.

\subsection{Neutral states}\label{N}

The positive energy eigenstates have wave functions $\phi_{\vec{p}
s}^{(+)(n)}(\vec{r})= e^{i \vec{p}.\vec{r}}\,u_{\vec{p} s}$, with
\begin{eqnarray}
u_{\vec{p} s}= N_{\vec{p} s}
\left(
\begin{array}{c}
1 \\ \\\frac{-s\,(p_x + i p_y)\,p_z}{(\Delta+s\,m_n)\,(E_{\vec{p} s}+s\,\Delta-\kappa_n B)} \\ \\
\frac{p_z}{E_{\vec{p} s}+s\,\Delta-\kappa_n B} \\ \\\frac{s\,(p_x + i p_y)}{\Delta+s\,m_n} \\
\end{array}
\right) \label{Spinorn}
\end{eqnarray}
and
\begin{eqnarray}
E_{\vec{p} s}&=&\sqrt{p_z^2+(\Delta-s\,\kappa_n B)^2}\\ \nonumber \\
\Delta&=&\sqrt{m_n^2+p^2_x+p^2_y}\\ \nonumber \\N_{\vec{p}
s}^2&=&\frac{1}{4\,(2 \pi)^3} \frac{(\Delta+s\,m_n)\,(E_{\vec{p}
s}+s\,\Delta-\kappa_n B)} {m_n\,(\Delta-s\,\kappa_n B)}.
\end{eqnarray}
On the other hand, the antiparticle states, of energy $-E_{\vec{p}
s}$, are $\phi_{\vec{p} s}^{(-) (n)}(\vec{r})=
e^{-i\vec{p}.\vec{r}}\,v_{\vec{p} s}$ with
\begin{eqnarray}
v_{\vec{p} s}=N_{\vec{p} s} \left(
\begin{array}{c}
\frac{p_z}{E_{\vec{p} s}+s\,\Delta-\kappa_n B} \\ \\
\frac{s\,(p_x + i p_y)}{\Delta+s\,m_n} \\ \\
1 \\ \\ \frac{- s\,(p_x + i p_y)\,p_z}{(\Delta+s\,m_n)\,(E_{\vec{p} s}+s\,\Delta-\kappa_n B)} \\
\end{array}
\right)
\end{eqnarray}
\\
Similarly to the previous case, these eigenstates are normalized
according to
\begin{eqnarray}
<\bar{\phi}_{\vec{p}\,' s}^{(\pm) (n)}|\phi_{\vec{p} s}^{(\pm)
(n)}>&=&\pm \delta^3(\vec{p}\,'-\vec{p})
\end{eqnarray}
and therefore satisfy the covariant orthogonal conditions
\begin{eqnarray}
<\phi_{\vec{p}\,' s'}^{(\pm) (n) \dag}|\phi_{\vec{p} s}^{(\pm)
(n)}> &=& \frac{E_{\vec{p}
s}\,\Delta}{m_n\,(\Delta-s\,\kappa_n\,B)}\,
\delta_{s' s}\,\delta^3(\vec{p}\,'-\vec{p})\nonumber \\
<\phi_{\vec{p}\,' s'}^{(+) (n)\dag}|\phi_{\vec{p} s}^{(-) (n)}>&=&
<\phi_{\vec{p}\,' s'}^{(-) (n) \dag}|\phi_{\vec{p} s}^{(+)(n)}>=0
\label{CNEUTRON}\end{eqnarray}

\section{Dirac fields and Green functions} \label{GreenP&N}

We propose an expansion of the fields in terms of creation and
destruction operators for states with the quantum numbers specified
in the previous section. Hence, the coefficients in this expansion
correspond to the wave functions previously described
\cite{RITUS}.\\
Thus, we obtain for the charged field
\begin{eqnarray}\begin{array}{l}
\Psi^{(p)}(t,\vec{r})=\int dp_y\,dp_z
\sqrt{\frac{m_p-\kappa_p\,B}{E_0}}\quad. \\ \\ \left[ e^{-i
E_0\,t}\,e^{i(p_y y+p_z z)}\,e^{-\xi^2/2}\, u_{0
p_z}(\xi)\,a^{(p)}_{0 p_y p_z} + e^{i E_0\,t}\,e^{-i(p_y y+p_z
z)}\,e^{-\eta^2/2}\,
v_{0 p_z}(\eta)\,d^{\dag(p)}_{0 p_y p_z} \right] \\ \\
+ \sum_{n=1}\sum_{s=\pm1}\int dp_y\,dp_z \sqrt{\frac{m_p\,(\Delta_n-s\,\kappa_p\,B)}{E_{n s}\,\Delta_n}}\quad. \\ \\
\left[
e^{-i E_{n s}\,t}\,e^{i(p_y y+p_z z)}\,e^{-\xi^2/2}\,
u_{n s p_z}(\xi)\,a^{(p)}_{n s p_y p_z} +
e^{i E_{n s}\,t}\,e^{-i(p_y y+p_z z)}\,e^{-\eta^2/2}\,
v_{n s p_z}(\eta)\,d^{\dag(p)}_{n s p_y p_z} \right] \\ \\
\end{array}\label{FPROTON}\end{eqnarray}
and for the neutral field
\begin{eqnarray}\begin{array}{l}
\Psi^{(n)}(t,\vec{r})=\sum_{s=\pm1} \int d\vec{p} \,
\sqrt{\frac{m_n\,(\Delta-s\,\kappa_n\,B)}{E_{\vec{p}
s}\,\Delta}}\left[ e^{-i E_{\vec{p} s}\,t}\,e^{i
\vec{p}.\vec{r}}\, u_{\vec{p} s}\,a^{(n)}_{\vec{p} s} + e^{i
E_{\vec{p} s}\,t}\,e^{-i\vec{p}.\vec{r}}\, v_{\vec{p}
s}\,d^{\dag(n)}_{\vec{p} s} \right]
\end{array}\label{FNEUTRON}\end{eqnarray}
where we have introduced an appropriate measure of integration in
the phase space \cite{ITZYKSON}. Written in this form, these expressions clearly
reduce to the more familiar ones when $\kappa_b=0$.

These fields satisfy the anti-commutation relations
\begin{eqnarray}\begin{array}{l}
\{\Psi^{\dag (b)}_\alpha(t,\vec{r}\,'),\Psi^{(b)}_\beta(t,\vec{r})\}=
\delta_{\alpha \beta}\,\delta^3(\vec{r}\,'-\vec{r})
\end{array}\label{CFIELD}\end{eqnarray}
if the standard anti-commutation relations are assumed for the creation
and annihilation operators, {\it i.e.}
\begin{eqnarray}\begin{array}{l}
\{a^{\dag(b)}_j,a^{(b)}_{j'}\}=\{d^{\dag(b)}_j,d^{(b)}_{j'}\}=\delta_{j j'}\\
\{d^{\dag(b)}_j,a^{(b)}_{j'}\}=\{a^{\dag(b)}_j,d^{(b)}_{j'}\}=0\\
\end{array}\end{eqnarray}
where the indices  $j, j'$ stand for a full set of quantum numbers, either discrete or continuum.\\
These fields are used to evaluate the in-medium causal propagator
\begin{eqnarray}\begin{array}{l}
i G^{(b)}_{\alpha \beta}(t',\vec{r}\,',t,\vec{r})=
<T[\Psi^{(b)}_\alpha(t',\vec{r}\,') \bar{\Psi}^{(b)}_\beta(t,\vec{r})]>= \\
\Theta(t'-t) <\Psi^{(b)}_\alpha(t',\vec{r}\,')
\bar{\Psi}^{(b)}_\beta(t,\vec{r})> - \Theta(t-t')
<\bar{\Psi}^{(b)}_\beta(t,\vec{r})
\Psi^{(b)}_\alpha(t',\vec{r}\,') > \label{Green}
\end{array}\end{eqnarray}
where $\Theta$ denotes the Heaviside step function.
Here the angular brackets must be regarded as an statistical mean
value, as obtained for instance, by evaluating the trace with the
density matrix of the system. The same average acting on the
products of a pair of creation and/or destruction operators
produce the well known results \cite{FETTER}
\begin{eqnarray}
<a^{(b)}_{j'} a^{\dag(b)}_j>&=&\delta_{j j'}-<a^{\dag(b)}_j a^{(b)}_{j'}> \label{ExpVal1}\\
<d^{(b)}_{j'} d^{\dag(b)}_j>&=&\delta_{j j'}-<d^{\dag(b)}_j d^{(b)}_{j'}>\\
<a^{\dag(b)}_j a^{(b)}_{j'}>&=&\delta_{j j'}\,n_F(T,E^{(b)}_j)\\
<d^{\dag(b)}_j d^{(b)}_{j'}>&=&\delta_{j j'}\,n_F(T,-E^{(b)}_j)
\label{ExpVal2}\
\end{eqnarray}
where $n_F$ denotes the Fermi occupation number
\begin{equation}
n_F(T,p_0)=\frac{\Theta(p_0)}{1+e^{(p_0-\mu_b)/T}} +
\frac{\Theta(-p_0)}{1+e^{-(p_0-\mu_b)/T}} \label{OCCNUMB}
\end{equation}
at temperature $T$ and chemical potential $\mu_b$ associated with
the conservation of the baryonic number.\\
The remaining combination of pairs have null expectation values.\\
The following expansions of the direct product of the spinors
(\ref{Spinorp}),(\ref{Spinorp0}),(\ref{Spinorn}) are particularly
useful
\begin{equation}
u_{0 p_z}(\xi)\otimes \bar{u}_{0
p_z}(\xi')=\frac{1}{(2\pi)^2}\frac{\sqrt{q B/\pi}}{4 (m_p-\kappa_p\,
B)} (E_0 \gamma_0-p_z \gamma_3+m_p-\kappa_p\, B) (1+i \gamma_1
\gamma_2)\label{ProDir1}
\end{equation}
\begin{multline}
u_{n s p_z}(\xi)\otimes \bar{u}_{n s
p_z}(\xi')=\frac{1}{(2\pi)^2}\frac{\sqrt{q
B/\pi}(\Delta_n+s\,m_p)}{2^{n+3} n! \;m_p
(\Delta_n-s\,\kappa_p\,B)}\Big [ H_n(\xi) (E_{n s} \gamma_0-p_z
\gamma_3 + s\,\Delta_n-\kappa_p\,B)  \\+ i\, \frac{m_p-s\,
\Delta_n}{\sqrt{q B}} H_{n-1}(\xi) (E_{n s} \gamma_0-p_z \gamma_3
-s\,\Delta_n+\kappa_p\,B) \gamma_1 \Big ](1+i\,\gamma_1
\gamma_2)\\ \left[H_n(\xi')+i\, \frac{m_p-s\, \Delta_n}{\sqrt{q
B}} H_{n-1}(\xi') \gamma_1\right]
\end{multline}
\begin{multline}
u_{\vec{p} s}\otimes \bar{u}_{\vec{p} s}=
\frac{1}{(2\pi)^3}\frac{i\, s\, \gamma^1 \gamma^2}{4 m_n
(\Delta-\kappa_n s B)}\left[E_{\vec{p} s} \gamma_0-p_z
\gamma_3+(s \Delta-\kappa_n\, B) \,i \gamma_1 \gamma_2 \right] \\
\left( -p_x \gamma_1-p_y \gamma_2+m_n+ s \Delta i \gamma_1
\gamma_2\right)\label{ProDir2}
\end{multline}
A similar result can be obtained for the case of antiparticles.

In order to evaluate (\ref{Green}) we insert the expansions
(\ref{FPROTON}) or (\ref{FNEUTRON}), and we use the expectation
values (\ref{ExpVal1})-(\ref{ExpVal2}). With the aim of unifying
the contributions coming from particles and antiparticles, we
apply the following relations
\begin{eqnarray}
\frac{i}{2 \pi} \int_{-\infty}^\infty dp_0 \frac{f(p_0) e^{-i p_0
(t'-t)}}{p_0^2-E^2+i\epsilon}&=& \Theta(t'-t) \frac{f(E) e^{-i E
(t'-t)}}{2 E} + \Theta(t-t') \frac{f(-E) e^{i E (t'-t)}}{2 E}
\nonumber \\
n_F(T,\pm E)&=&2 \int_{-\infty}^\infty dp_0  |p_0|\; \Theta (\pm
p_0) \;\delta(p_0^2-E^2)\; n_F(T,p_0) \nonumber
\end{eqnarray}
($E>0$). Finally we make use of Eqs.
(\ref{ProDir1})-(\ref{ProDir2}) and similar relations for
antiparticles, to obtain the following results

\begin{eqnarray}
G^{(p)}_{\alpha \beta}(t',\vec{r}\,',t,\vec{r})&=& \frac{1}{2}
\sqrt{\frac{q B}{\pi}} \int \frac{dp_0\,dp_y\,dp_z}{(2 \pi)^3}
e^{-i p_0\,(t'-t)}\,
e^{i[p_y (y'-y)+p_z (z'-z)]}\,e^{-(\xi'^2+\xi^2)/2}
\nonumber \\
&&\! \! \! \! \! \! \bigg[ \Lambda^{0 (p)}_{\alpha \beta} \;
\Xi(T,E_0) + \sum_{n=1}\sum_{s=\pm1}
\frac{\Delta_n+s\,m_p}{2^{n+1}\,n!\,\Delta_n} \Lambda^{n s
(p)}_{\alpha \beta}(\xi',\xi) \; \Xi(T,E_{n s}) \bigg]
\label{GP}\end{eqnarray}
 for charged particles, where
\begin{eqnarray}
\Lambda^{0 (p)}&=&\left(
p_0\gamma^0-p_z\gamma^3+m_p-\kappa_p\,B\right)
\left(1+i\gamma^1\gamma^2\right) \\
\Lambda^{n s (p)}&=&\Big[(p_0\gamma^0-p_z\gamma^3+s
\Delta_n-\kappa_p B) H_n(\xi')+i  \frac{m_p - s \Delta_n}{\sqrt{q
B}} (p_0\gamma^0-p_z\gamma^3-s \Delta_n+\kappa_p B) \gamma^1
H_{n-1}(\xi')\Big ] \nonumber \\
&& \left[(1+i\gamma^1\gamma^2)
H_n(\xi)+i \frac{m_p - s \Delta_n}{\sqrt{q B}}
\gamma^1(1-i\gamma^1\gamma^2)
H_{n-1}(\xi)\right] \\
\Xi(T,E)&=&\frac{1}{p_0^2-E^2+i\epsilon}+2
\pi\,i\,n_F(T,p_0)\,\delta(p_0^2-E^2)
 \label{TFD} \end{eqnarray}
and $\xi'=(-p_y + q B x')/\sqrt{q B}$. \\
In addition we have
\begin{eqnarray}\begin{array}{l}
G^{(n)}_{\alpha \beta}(t',\vec{r}\,',t,\vec{r})= \sum_{s=\pm1}
\int \frac{d^4p}{(2 \pi)^4} e^{-i p^\mu\,(x'_\mu-x_\mu)}
\Lambda^{s \, (n)}_{\alpha \beta} \; \Xi(T,E_{\vec{p} s})
\end{array}\label{GN}\end{eqnarray}
 for neutral particles, where
\begin{eqnarray}\begin{array}{l}
\Lambda^{s \,(n)}=\frac{-i s\; \gamma^1 \gamma^2}{2 \Delta}\left[
p_0\gamma^0-p_z\gamma^3+ i \gamma^1 \gamma^2 (s \Delta-\kappa_n
B)\right] \left( p_x\gamma^1+p_y\gamma^2-m_n- i s \Delta \gamma^1
\gamma^2\right)
\end{array}\nonumber \end{eqnarray}
eqs. (\ref{GP}) and (\ref{GN}) resume the main findings of this
work.  Of course, these Green functions satisfy the differential
equation
\[
\left(i \gamma^\mu D_\mu-m_b+i
\gamma^1\gamma^2\,\kappa_b\,B\right) G^{(b)}(x,x')=\delta^4(x-x')
\nonumber
\]
where  $D_\mu=\partial_\mu + i q_b\,A_\mu $.

It is a well known fact that real time formulations of the thermal
field theory \cite{CHOU,LANDSMAN}, like Schwinger-Keldysh theory
or Thermo Field Dynamics (TFD), needs to duplicate the degrees of
freedom in order to keep the formalism and procedures of the usual
field theory. In TFD for instance, to each physical field
$\varphi^{(1)}(x)$ there corresponds a dual partner
$\varphi^{(2)}(x)$, and they are related by the so called tilde
conjugation operation. As a consequence, there is a $2\times 2$
matrix
 associated to the product of two fields. This is also the case for the one-particle propagators
$i\, G^{a b}(x,x')=<T \varphi^{(a)}(x) \varphi^{(b) }(x')>$, and
the corresponding self-energies. Within this context the results
shown in Eqs.(\ref{GP}) and (\ref{GN}) correspond to the component
(1,1) of the TFD representation. However, it suffices to treat the
MFA at zero temperature to be developed in the next section.\\
To evaluate higher order corrections to the finite temperature
self energy, the full dependence on the thermal degrees of freedom
must be taken into account. This means that for a given
perturbative diagram, for each internal line there corresponds a
$2\times 2$ propagator and a sum over the thermal index $c=1,2$
should be included for each internal vertex \cite{MATSUMOTO}. \\
A resume of the Feynman graph rules in TFD for the QHD model are
given in Ref.\cite{SAITO}.\\
Within the quasi-particle scheme described at the beginning of
this section, it is no difficult to evaluate some thermal
expectation values required to complete the TFD propagator. In
practice,  Eq.(\ref{TFD}) must be replaced by the following matrix
\begin{eqnarray}
\Xi(T,E)=\left( \begin{array}{cc}
\frac{1}{p_0^2-E^2+i\varepsilon}&0\\
0&\frac{1}{p_0^2-E^2-i\varepsilon}
\end{array}\right)+2 \pi i \delta(p_0^2-E^2)\left( \begin{array}{cc}
n_F(T,p_0)&\bar{n}_F(T,p_0)\\
\bar{n}_F(T,p_0)&-n_F(T,p_0)
\end{array}\right) \label{OCCNUMB2}
\end{eqnarray}
with $n_F$ as in Eq.(\ref{OCCNUMB}) and
\begin{equation}
\bar{n}_F(T,p_0)=\frac{e^{(p_0-\mu_b)/2 T}}{1+e^{(p_0-\mu_b)/T}}
\Theta(p_0)- \frac{e^{-(p_0-\mu_b)/2
T}}{1+e^{-(p_0-\mu_b)/T}}\Theta(-p_0) \nonumber
\end{equation}
It must be noticed that at zero temperature the matrix in
Eq.(\ref{OCCNUMB2}) becomes diagonal, that is, the thermal degrees
of freedom are decoupled.

In the next section we study the coherence of these results, by
comparing some simple calculations with well established facts of
the QHD formalism.

\section{Mean values of nuclear matter densities}\label{QHD}

In order to keep the simplicity of the discussion, we have reduced
the problem to its bare minimum. Up to this point we have regarded
the nucleon as a non-interacting particle. Now, we include the
strong interaction between the nucleons and its environment in the
MFA of the QHD model. Within this scheme, the lightest mesons dress
the nucleon giving rise to a quasi-particle picture. The net effect
is to modify the mass of the nucleons by $m_b^*=m_b-g_\sigma \,
\sigma_0$, as a consequence the modified quantities $\Delta_n^*, \,\Delta^*$ must be introduced.
The single particle spectrum becomes
$\epsilon_{b \,
j}=E_j+g_\omega\,\omega_0$, where $\sigma_0,\omega_0$ are the uniform in-medium expectation values
of the meson fields $f_0(500)$ and  $\omega(782)$, respectively \cite{SW}.
\\
We can include these modifications into the propagators (\ref{GP})
and (\ref{GN}) by assuming that the masses $m_p^*$ and $m_n^*$
represent the in-medium values described above, while the change in
the energy spectrum can be handled by replacing the chemical
potentials $\mu_b$, which are implicit in the Fermi distribution
functions $n_F$, by the effective ones
$\bar{\mu}_b=\mu_b-g_\omega\,\omega_0$.\\
In the following we check the validity of the results presented in
the previous section by comparing our calculations for the nuclear
scalar, baryon and energy densities, with those presented in
previous calculations. In the spirit of the MFA we neglect the
divergent contributions coming from the Dirac sea.
We start examining the average mean field scalar $\rho^{(b)}_s$ and baryon $\rho^{(b)}$ densities, using the general definitions
\begin{eqnarray}
\rho^{(b)}_s(t,\vec{r})=&&<\bar{\Psi}^{(b)}(t,\vec{r})\,\Psi^{(b)}(t,\vec{r})>\nonumber \\
=&&-i\;\lim_{(t'\rightarrow t^+,\vec{r}\,'\rightarrow\vec{r})\;\;}\,Tr\{G^{(b)}(t,\vec{r},t',\vec{r}\,')\}\\
\nonumber \\
\rho^{(b)}(t,\vec{r})=&&<\bar{\Psi}^{(b)}(t,\vec{r})\,
\gamma^0\,\Psi^{(b)}(t,\vec{r})>\nonumber \\
=&&-i\;\lim_{(t'\rightarrow
t^+,\vec{r}\,'\rightarrow\vec{r})\;\;}\,Tr\{\gamma^0\,G^{(b)}(t,\vec{r},t',\vec{r}\,')\}
\end{eqnarray}
For the case of protons the results are
\begin{eqnarray}
\rho^{(p)}_s=&&\frac{q B}{2 \pi^2} \int_0^\infty dp_z \{\frac{(m_p^*-\kappa_p\,B)}{\epsilon_0}[n_F(T,\epsilon_0)+n_F(T,-\epsilon_0)]\nonumber \\
+&&m_p^*\,\sum_{n=1}\sum_{s=\pm1} \frac{(\Delta^*_n-s\,\kappa_p
B)}{\epsilon_{n s}\,\Delta^*_n}[n_F(T,\epsilon_{n
s})+n_F(T,-\epsilon_{n s})] \}
\\ \nonumber \\
\rho^{(p)}=&&\frac{q B}{2 \pi^2} \int_0^\infty dp_z
\{[n_F(T,\epsilon_0)-n_F(T,-\epsilon_0)]\nonumber \\+&&
\sum_{n=1}\sum_{s=\pm1} [n_F(T,\epsilon_{n s})-n_F(T,-\epsilon_{n
s})] \}
\end{eqnarray}
Similarly for neutrons
\begin{eqnarray}
\rho^{(n)}_s=&&\int \frac{d^3p}{(2 \pi)^3} \sum_{s=\pm1}
\frac{(\Delta^*-s\,\kappa_n B)}{\epsilon_{\vec{p}
s}\,\Delta^*}[n_F(T,\epsilon_{\vec{p} s})+n_F(T,-\epsilon_{\vec{p}
s})]
\\ \nonumber \\
\rho^{(n)}=&&\int \frac{d^3p}{(2 \pi)^3} \sum_{s=\pm1}
[n_F(T,\epsilon_{\vec{p} s})-n_F(T,-\epsilon_{\vec{p} s})]
\end{eqnarray}
As the temperature tends to zero, $T\rightarrow0$, we have
$n_F(T,-\epsilon^{(b)}) \rightarrow 0$,
$n_F(T,\epsilon^{(b)})\rightarrow\Theta(\bar{\mu}_b-\epsilon^{(b)})$.
The last condition defines the Fermi momentum for protons and
neutrons. For the first case we have
$p^{(p)}_{Fs}=\sqrt{\bar{\mu}_p^2-(\Delta^*_{n \max}-s \kappa_p
B)^2}$, with the highest occupied Landau level given by the
condition $|\Delta^*_{n \max}-s \kappa_p B| = \bar{\mu}_p$. While
for neutrons the Fermi surface is defined by
$p^{(n)}_{Fs}=\sqrt{\bar{\mu}_n^2-(\Delta^*_{\max}-s \kappa_n
B)^2}$ along the z-axis and by $|\Delta^*_{\max}-s \kappa_n B| =
\bar{\mu}_n$ in the orthogonal plane.

The baryonic contribution to the energy density $\varepsilon^{(b)}$
arises from the mean field value of the Hamiltonian density operator
$\mathcal{H}^{(b)}(t,\vec{r})=\bar{\Psi}^{(b)}(t,\vec{r})\,(i
\gamma^0 \partial/\partial t)\,\Psi^{(b)}(t,\vec{r})$, {\it i.e.}
\begin{eqnarray}
\varepsilon^{(b)}=<\mathcal{H}^{(b)}>=-i\;\lim_{(t'\rightarrow
t^+,\vec{r}\,'\rightarrow\vec{r})\;\;}\,Tr\{(i \gamma^0
\partial/\partial t)\,G^{(b)}(t,\vec{r},t',\vec{r}\,')\}
\end{eqnarray}
Thus, we have
\begin{eqnarray}
\varepsilon^{(p)}=&&\frac{q B}{2 \pi^2} \int_0^\infty dp_z \{\epsilon_0\,[n_F(T,\epsilon_0)+n_F(T,-\epsilon_0)]\nonumber \\
+&& \sum_{n=1}\sum_{s=\pm1} \epsilon_{n s}\,[n_F(T,\epsilon_{n s})+n_F(T,-\epsilon_{n s})] \}\\ \nonumber \\
\varepsilon^{(n)}=&&\int \frac{d^3p}{(2 \pi)^3} \sum_{s=\pm1}
\epsilon_{\vec{p} s}\, [n_F(T,\epsilon_{\vec{p}
s})+n_F(T,-\epsilon_{\vec{p} s})]
\end{eqnarray}
By taking the limit $T\rightarrow0$ of these results we find a
complete agreement with the calculations of \cite{BRODERICK}.

\section{Conclusions}
In this work we have evaluated the covariant propagator for nucleons
in the presence of a strong magnetic field. We have extended
previous results by including the intrinsic magnetic moment and the
effects of finite density and temperature. We have performed a
detailed derivation with a clear statement of the notation used, a
fact that was lacking in the literature.\\
 Furthermore, our results have been interpreted in
the context of the Thermo Field Dynamics, and definite expressions
for the thermal propagator as $2 \times 2$ matrix have been
presented.\\
 We have performed some simple calculations which show the
coherence with previous results. We consider that the relevance of
our findings lies in the fact that it allows the evaluation of in
medium nuclear processes by including the full effects of an
external magnetic field, and further corrections can be included
by using
the diagrammatic techniques of relativistic Quantum Field Theory.\\
Self consistent calculations of corrections to the
propagation of hadrons in matter under strong magnetic fields are
now in progress.

\section{Acknowledgements}
This work has been partially supported by the CONICET of Argentina
under grant PIP 112-2008-01-00282.

\end{document}